\documentclass[aps,amsmath,onecolumn,amssymb,nofootinbib,floatfix,showpacs]{revtex4}
\usepackage{graphicx}
\usepackage{amsmath}
\usepackage{amssymb}
\usepackage{latexsym}
\usepackage{dcolumn}
\usepackage{bm}
\usepackage{hyperref} 
\usepackage{url}
\tighten

\begin{document}
\title{The linear sigma model at a finite isospin chemical potential}
\author{Hong Mao$^1$}
\email{maohong@hp.ccast.ac.cn}
\author{Nicholas Petropoulos$^2$}
\author{Song Shu$^3$}
\author{ Wei-Qin Zhao$^{1,4}$}
\address{1. CCAST(World Laboratory), P.O. Box 8730, Beijing 100080,
China  \\
2. Department of Electronics, Technological Educational Institute of Lamia, Lamia 35100, Greece\\
3.Faculty of Physics and Electronic Technology, Hubei University,
Wuhan 430062, China\\
4.Institute of High Energy Physics, Chinese Academy of Sciences,
Beijing 100049, China}

\begin{abstract}
The effect of finite isospin chemical potential to the effective
masses of the mesons at finite temperature is investigated  in the
framework of  the $O(4)$ linear sigma model with explicit chiral
symmetry breaking. We present a mechanism to include the isospin
chemical potential in the model. By using the
Cornwall-Jackiw-Tomboulis method of composite operators, we obtain a
set gap equations for the effective masses of the mesons and get the
numerical results in the Hartree approximation. We find that the
introduction of the chemical potential only affects the mass of the
charged pions and sigma, while there is almost NO effects on the
mass of neutral pions.
\end{abstract}


\maketitle

\section{\label{Section:Intro}Introduction}

Nowadays, it is widely believed that QCD is the most promising
theory to explain the strong interactions. However, due the property
of confinement and asymptotic freedom we can accept and study the
quarks as free only at very high energies. This much expected phase,
which is called Quark Gluon Plasma (QGP), is under investigation in
RHIC experiments and there is strong evidence that it has been
observed. According to Heinz \cite{Heinz:2004pj} it seems that
elliptic flow at high enough collision energies is a QGP signature.
The recent results from RHIC experiments have attracted many
attempts for theoretical models to be tested\cite{Rischke:2003mt}.
Recently there are also a lot of attempts to study the phases of QCD
from first principles using lattice techniques at finite baryon
density and chemical isospin potential. There is, however, an
alternative approach which applies only in the low energy regime.
Following this path, it is necessary to use a variety of effective
phenomenological models and QCD--like theories in order to mimic
some of the features of QCD.

A quite popular model, which fulfils these requirements, exhibiting
many of the symmetries observed in QCD, is the linear sigma model.
The model was first introduced in the 1960s as a model for
pion--nucleon interactions \cite{Gell-Mann:1960np} and has attracted
much attention recently, especially in studies involving disoriented
chiral condensates
\cite{Rajagopal:1992qz,Rajagopal:1993ah,Kuraev:2003tr}.  The linear
sigma model with a $U(N_f)\times U(N_f)$ symmetry and two to four
quark flavors have been studied in the Hartree-Fock or Hartree
approximation\cite{Petropoulos:1998gt,Lenaghan:2000ey,Roder:2003uz,Petropoulos:2004bt}
within the Cornwall--Jackiw--Tomboulis (CJT)
formalism\cite{Amelino-Camelia:1992nc}, and the large-N
approximation of the linear sigma model has been investigated by
using the same formlism in
Refs\cite{Amelino-Camelia:1997dd,Lenaghan:1999si,Nemoto:1999qf}.
Within the same framework, Shu and Li \cite{Shu:2005bj}have studied
Bose-Einstein condensation and the chiral phase transition in the
chiral limit, and they introduced the chemical potential through the
third component of isospin charge. In this discussion, we apply a
mechanism \cite{Kogut:1999iv,Kogut:2000ek,Son:2000xc} to include the
isospin chemical potential in the linear sigma model, then by using
CJT method, we study the chiral phase transition at finite isospin
chemical potential in explicit chiral symmetry breaking.

The organization of this paper is as follows. In the next section we
present the basics about the linear sigma model and how this model
incorporates some of the basic features of QCD, then we introduce
the isospin chemical potential in the model. In
Section~\ref{Section:Veff} we calculate the effective potential of
the model by using the Cornwall--Jackiw--Tomboulis method of
composite operators and obtain a set of gap equations for the
thermal effective mass of the pions. The solution of these gap
equations is presented in Section~\ref{Section:Numerical}. In the
last section we present our conclusions.

\section{The \label{Section:Model}Model}

Consider an idealisation of QCD with two species of massless quarks
$u$ and $d$. The Lagrangian of strong interaction physics is
invariant under $SU(2)_L\times SU(2)_R$ chiral transformation
\begin{equation}
\Psi _{L,R}\rightarrow \exp (-i\vec{\theta}_{L,R}\cdot
\vec{\tau})\Psi _{L,R},
\end{equation}
where $\Psi _{L,R}=(u,d)_{L,R}$. However this chiral symmetry does
not appear in the low energy particle spectrum since it is
spontaneously broken. Consequently, three Goldstone bosons, the
pions, appear and the (constituent) quarks become massive. At low
energy, the spontaneous breaking of chiral symmetry can be described
by an effective theory, the linear sigma model, which involves the
massless pions $\vec{\pi}$ and a massive $\sigma $ particle. Here,
the $\sigma$ field can be used to represent the quark condensate
which is considered as the order parameter of the chiral phase
transition. This  choice is a natural one since both exhibit the
same behaviour under chiral transformations. The pions are very
light particles and can be considered approximately as massless
Goldstone bosons.

There are different ways to write down the Lagrangian of the linear
sigma model and each with  certain advantages and disadvantages. One
version which demonstrates the chiral properties of this Lagrangian
in a straightforward way involves the introduction of a matrix field
$\Phi$ defined as
\begin{equation}
\Phi =\sigma \frac{\tau ^0}2+i\vec{\pi}\cdot \frac{\vec{\tau}}2~,
\end{equation}
where $\tau ^0$ is the unity matrix and
$\vec{\tau}=(\tau_1,\tau_2,\tau_3)$ are the Pauli matrix. This
matrix field satisfies  the normalization condition $Tr(\tau ^a\tau
^b)=2\delta ^{ab}$. Under $SU(2)_L\times SU(2)_R$ chiral
transformations, $\Phi $ transforms as
\begin{equation}
\Phi \rightarrow L^{+}\Phi R~.
\end{equation}
Using this parametrisation the renormalizable effective Lagrangian
of the linear sigma model can be written as  \cite{Itzykson:1980rh}
\cite{Donoghue:1992dd}
\begin{equation}\label{Eq:Lag0}
{\mathcal{L}}_0={\mathcal{L}}_\Phi +{\mathcal{L}}_q~,
\end{equation}
where
\begin{equation}\label{Eq:Lag1}
{\mathcal{L}}_\Phi =\textrm{Tr}[(\partial _\mu \Phi )^{+}(\partial
^\mu \Phi )]
-m^2\textrm{Tr}(\Phi^{+}\Phi)-\frac{\lambda}{6}[\textrm{Tr}(\Phi
^{+}\Phi)]^2~
\end{equation}

\begin{equation}\label{Eq:Lag2}
{\mathcal{L}}_q=\overline{\Psi}_Li\gamma ^\mu \partial _\mu \Psi_L
+\overline{\Psi}_R i\gamma ^\mu \partial_\mu
\Psi_R-2g\overline{\Psi}_L \Phi \Psi_R+h.c..
\end{equation}
When chiral symmetry is breaking, the field $\sigma$ acquires a
non--vanishing vacuum expectation value, which breaks $SU(2)_L\times
SU(2)_R$ down to $SU(2)_{L+R}$. It results in a massive sigma
particle $\sigma $ and three massless Goldstone bosons $\vec{\pi}$,
as well as giving a mass $m_q=gf_\pi$ to the constituent quarks.

Now let us consider the Lagrangian of the linear sigma model for two
massless quarks of flavors $u$, $d$ with different chemical
potential $\mu_u$ and $\mu_d$,
\begin{eqnarray} \label{Eq:Lag3}
{\mathcal{L}}'  =  {\mathcal{L}}_{0}+{\mathcal{L}}_{\mu}
               =  {\mathcal{L}}_{\Phi}+{\mathcal{L}}_{q}+\Psi_{L}^\dag B\Psi_{L}^{}+\Psi_{R}^\dag B\Psi_{R}^{},
\end{eqnarray}
where $B=\left(\begin{array}{c} \mu_u\\0
\end{array} \begin{array}{c} 0\\\mu_d
\end{array}\right)$
is the matrix of chemical potentials. The term ${\mathcal{L}}_{\mu}$
can be expressed either by the variables $\mu_u$, $\mu_d$ or by the
two combinations $\mu_B=\frac{1}{2}(\mu_u+\mu_d)$ and
$\mu_I=\frac{1}{2}(\mu_u-\mu_d)$, which couple to the baryon charge
density and the third component of isospin respectively. Then we
have
\begin{equation}\label{Eq:Lag4}
{\mathcal{L}}_{\mu}=\mu_B(\Psi^{\dag}_{L}\Psi_{L}^{}+\Psi^{\dag}_{R}\Psi_{R}^{})+
\mu_I(\Psi^{\dag}_{L} \tau_3 \Psi_{L}^{}+\Psi^{\dag}_{R} \tau_3 \Psi_{R}^{})~.
\end{equation}
Both ${\mathcal{L}}_{\Phi}$ and ${\mathcal{L}}_{q}$ are $SU(2)_{L}
\times SU(2)_{R}$ invariant, but the symmetry is reduced by the
baryon charge density term
$\mu_B(\Psi^{\dag}_{L}\Psi_{L}^{}+\Psi^{\dag}_{R}\Psi_{R}^{})$ to
$SU(2)_{L+R}$ and further reduced by the isospin term
$\mu_I(\Psi^{\dag}_{L} \tau_3 \Psi_{L}^{}+\Psi^{\dag}_{R} \tau_3
\Psi_{R}^{})$ to $U(1)_{L+R}^{}$. In what follows, we consider an
idealized case where $\mu_{I}$ is nonzero while $\mu_B=0$. A similar
choice is adopted in Refs.\cite{Son:2000xc}\cite{Barducci:2004tt}.
According to their analysis such an approximation is justified by
the fact that we are dealing with the dynamics of strong
interactions. In any realistic setting, $\mu_I \ll \mu_B$. Such a
system is unstable due to weak decays which do not conserve isospin.
However, we can still imagine that all relatively slow electroweak
effects are turned off.

We note that the effects of the isospin chemical potential $\mu_{I}$
can be included to the Lagrangian of the linear sigma model without
additional phenomenological parameters by promoting $SU(2)_{L}
\times SU(2)_{R}$ to a local gauge symmetry and viewing $\mu_{I}$ as
the zeroth component of a gauge potential
\cite{Kogut:1999iv,Kogut:2000ek,Son:2000xc}. This means that we can
rewrite the terms ${\mathcal{L}}_{q}+{\mathcal{L}}_{\mu}$ as
\begin{equation} \label{Eq:Lag5}
{\mathcal{L}}_{q}'=\bar{\Psi}_{L}i
\gamma^{\mu}D_{\mu}\Psi_{L}+\bar{\Psi}_{R}i
\gamma^{\mu}D_{\mu}\Psi_{R}~,
\end{equation}
where $D_{\mu}=\partial_{\mu}-i \mu_{I}A_{\mu}$,
$A_{\mu}=A_{\mu}^aT^a$ and $T_a=\frac{\tau^a}{2} (a=1,2,3)$. Here
the gauge fields $A_{\mu}$ have the following properties:
$A^{3}_{0}={\mathbf{1}}, A^{1}_{0}=A^{2}_{0}=0$ and $A_{i}=0$ for
$i=1, 2, 3$. Then the term ${\mathcal{L}}_{q}'$ is invariant under
the $SU(2)_L\times SU(2)_R$ local transformations:
\begin{equation}
\Psi _{L,R}\rightarrow U\Psi _{L,R},U=\exp
(-i\vec{\theta}_{L,R}(x)\cdot \vec{T})~.
\end{equation}
However, the term ${\mathcal{L}}_{\Phi}$ is not invariant under
$SU(2)_L\times SU(2)_R$ local transformations. In order to ensure
the effective Lagrangian of the linear sigma model is also
invariant under this local symmetry we have to replace the
derivatives in Eq.~(\ref{Eq:Lag1}) by the covariant derivatives:
\begin{equation}\label{Eq:Lag6}
{\mathcal{L}}_\Phi=\textrm{Tr}[(D_\mu \Phi )^{+}(D_\mu \Phi
)]-m^2\textrm{Tr}(\Phi
^{+}\Phi)-\frac{\lambda}{6}(\textrm{Tr}(\Phi ^{+}\Phi))^2~,
\end{equation}
where the covariant derivatives are defined as:
\begin{eqnarray}
D_0\Phi &=&
\partial_{0}\Phi-i\mu_{I}(\frac{\tau^3}{2}\Phi-\Phi\frac{\tau^3}{2})
\\
D_{i}\Phi &=& \partial_{i}\Phi, (i=1,2,3)~.
\end{eqnarray}
Thus the gauge invariance fixes the way $\mu_{I}$ enters the
mesonic sector of the effective Lagrangian of the linear sigma
model.

In order to study the effects of the finite chemical potential
$\mu_{I}$ in the linear sigma model at finite temperature, we only
consider the term ${\mathcal{L}}_\Phi$ neglecting the fermion
fields. It is convenient to define the new fields:
\begin{eqnarray}
\pi^{\pm}= \frac{\pi^1 \pm i\pi^2}{\sqrt{2}}~.
\end{eqnarray}
The Lagrangian ${\mathcal{L}}_\Phi$ now can be expressed as
\begin{eqnarray}\label{Eq:Lag7}
{\mathcal{L}}=
\frac{1}{2}(\partial_{\mu}\sigma)(\partial^{\mu}\sigma)
+\frac{1}{2}(\partial_{\mu}\pi^0)(\partial^{\mu}\pi^0)+(D_{\mu}\pi)^{+}(D^{\mu}\pi)^{-}-V(\sigma,
\vec{\pi})~,
\end{eqnarray}
where the potential is
\begin{equation}
V(\sigma,
\vec{\pi})=\frac{m^2}{2}(\sigma^2+\vec{\pi}^2)+\frac{\lambda}{24}(\sigma^2+\vec{\pi}^2)^2~,
\end{equation}
and $(D_{0}\pi)^{\pm}=(\partial_{0} \pm i\mu_{I})\pi^{\pm}$,
$(D_{i}\pi)^{\pm}=\partial_{i}\pi^{\pm}$ for $(i=1,2,3)$.

The $SU(2)_L\times SU(2)_R$ symmetry of the linear sigma model is
explicitly broken if the potential $V(\sigma, \vec{\pi})$ is made
slightly asymmetric by adding to the basic Lagrangian in
Eq.~(\ref{Eq:Lag0}) a term of form
\begin{equation}
{\mathcal{L}}_b=\epsilon \sigma~.
\end{equation}
With this addition, the vector isospin $SU(2)$ symmetry remains
exact, but the axial $SU(2)$ transformation is no longer invariant.
The potential now has the form
\begin{equation} \label{Eq:Potential1}
V(\sigma,
\vec{\pi})=\frac{m^2}{2}(\sigma^2+\vec{\pi}^2)+\frac{\lambda}{24}(\sigma^2+\vec{\pi}^2)^2-\epsilon\sigma~.
\end{equation}
To the leading order in $\epsilon$, this shifts the minimum of the
potential to $v=f_{\pi}+\frac{3 \epsilon}{\lambda f_{\pi}^2}$, and
as a result the pions acquire mass
$m^2_{\pi}=\frac{\epsilon}{f_{\pi}}$.

At tree level and zero temperature the parameters of the Lagrangian
are fixed in a way that these masses agree with the observed value
of pion mass $m_{\pi}=138$ MeV and the most commonly accepted value
for sigma mass $m_{\sigma}=600$ MeV. Then, the coupling constant
$\lambda$ of the model can be related to zero temperature properties
of the pions and sigma through the expression
\begin{equation}
\lambda =\frac{3(m^2_{\sigma}-m^2_{\pi})}{f^2_{\pi}}~,
\end{equation}
where $f_\pi \approx 93$ MeV is the pion decay constant. The
negative mass parameter $m^2$ is introduced in order to ensure
spontaneous breaking of symmetry and its value is chosen to be
\begin{equation}
-m^2=(m^2_{\sigma}-3m^2_{\pi})/2>0~.
\end{equation}

\section{\label{Section:Veff}The CJT effective potential}

In this section we calculate the finite temperature effective
potential for the linear sigma model at finite isospin chemical
potential. In our calculations we use the imaginary time formalism,
which is also known as the Matsubara
formalism\cite{Dolan:1973qd,Kapusta:1989,LeBellac:1996,Das:1997gg}.
This means that in the case of Bosons, we have
\begin{eqnarray}
\int\frac{d^4k}{(2\pi)^4}f(k) \rightarrow \frac{1}{\beta}\sum_n
\int\frac{d^3\vec{k}}{(2\pi)^3}f(i\omega_n,\vec{k}) \nonumber
\equiv \int_{\beta}f(i\omega_n,\vec{k}),
\end{eqnarray}
where $\beta$ is the inverse temperature, $\beta=\frac{1}{k_BT}$,
and as usual Boltzmann's constant is taken as $k_B=1$, and
$\omega_n=2\pi nT$, $n=0, \pm1, \pm2, \pm3,\ldots$ For simplicity we
have introduced a subscript $\beta$ to denote integration and
summation over the Matsubara frequency sums.

The starting point is the Lagrangian given in Eq.~(\ref{Eq:Lag7})
with the choice of the asymmetric potential given in
Eq.~(\ref{Eq:Potential1}). By shifting the sigma field as
$\sigma\rightarrow \sigma+\phi$ the classical potential takes the
form
\begin{equation}
U(\phi)=\frac{1}{2}m^2\phi^2+\frac{\lambda}{24}\phi^4-\epsilon\phi.
\end{equation}
The tree-level sigma and pion propagators corresponding to the
above Lagrangian have the form
\begin{subequations}
\begin{eqnarray}
D_{\sigma}^{-1} &=& \omega_{n}^2+\vec{k}^2+m^2+\frac{1}{2}\lambda \phi^2, \\
D_0^{-1} &=&\omega_{n}^2+\vec{k}^2+m^2+\frac{1}{6}\lambda \phi^2, \\
D^{-1}_c &=& (\omega_{n}+
i\mu_{I})^2+\vec{k}^2+m^2+\frac{1}{6}\lambda \phi^2. \label{eq-pro3}
\end{eqnarray}
\end{subequations}
Eq.(\ref{eq-pro3}) represents the inverse propagator of $\pi_+$ and
$\pi_-$ due to the facts that $\omega_{n}+ i\mu_{I}$ and
$\omega_{n}-i\mu_{I}$ are equivalent in describing the propagators
of $\pi_+$ and $\pi_-$\cite{Shu:2005bj}.

The interaction Lagrangian which describes the vertices of the
shifted theory is given by
\begin{eqnarray}
{\mathcal{L}}_{int}=-\frac{\lambda}{12}\sigma^2
\vec{\pi}^2-\frac{\lambda}{24}\sigma^4-\frac{\lambda}{24}\vec{\pi}^4
-\frac{\lambda}{6}\phi\sigma\vec{\pi}^2-\frac{\lambda}{6}\phi\sigma^3~.
\end{eqnarray}
Here, we omit the constants and the terms linear in the $\sigma$
fields for simplicity. In our approximation we do not consider
interactions which are given by the last two terms in the above
Lagrangian.

\begin{figure}
\includegraphics[scale=0.8]{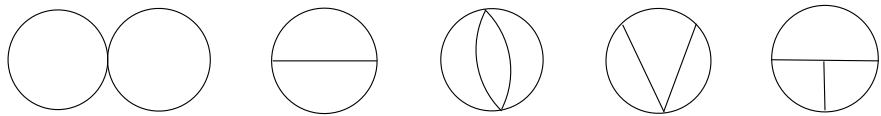}
\caption{\label{Fig:Fig1}Two--particle irreducible graphs which
contribute to the effective potential of a $\lambda\phi^4$ theory in
the CJT method up to the three-loop level. The solid line represents
the dressed propagator $G(x,y)$. There are two kinds of vertices: a
three-particle vertex and a four-particle vertex.}
\end{figure}

Then, for constant $\phi$, according to CJT
formalism \cite{Amelino-Camelia:1992nc} the finite temperature effective
potential for $\lambda\phi^4$ theory is given by
\begin{eqnarray}
V(\phi, G)&=& U(\phi)+\frac{1}{2}\int_{\beta}\ln
G^{-1}(\phi;k) \nonumber\\&&
+\frac{1}{2}\int_{\beta}[D^{-1}(\phi;k)G(\phi;k)-1]\nonumber\\&&
+V_2(\phi, G).
\end{eqnarray}
Here, $G(\phi;k)$ denotes the full propagator of the theory and we
have introduced a shorthand notation for the space integration and
summation over the bosonic Matsubara frequencies so the symbol
$\int_{\beta}$ stands for $T\sum_n\int d^3\vec{k}/(2\pi)^3$. The
last term $V_2(\phi, G)$ represents  the sum of all two and
higher--order loop two--particle irreducible vacuum graphs of the
theory with vertices given by ${\mathcal{L}}_{int}$ and propagators
set equal to $G(\phi;k)$. The diagrams contributing to $V_2(\phi,
G)$  for $\lambda\phi^4$ theory are shown in Fig.~\ref{Fig:Fig1}.
Evaluating the effective potential in Hartree approximation means
that one needs to take into account the "$\infty$" type diagram
only. In the case of $\lambda\phi^4$ theory this diagram is the
leading order in $V_2(\phi, G)$ in both the loop expansion and the
$1/N$ expansion. We adopt this approximation in the case of the
linear sigma model as well. The Hartree and large N approximations
at zero chemical potential have been studied previously in
Refs.\cite{Petropoulos:1998gt,Lenaghan:2000ey,Roder:2003uz,Petropoulos:2004bt,Amelino-Camelia:1997dd,Lenaghan:1999si,Nemoto:1999qf}.

\begin{figure}
\begin{center}
\includegraphics[scale=0.59]{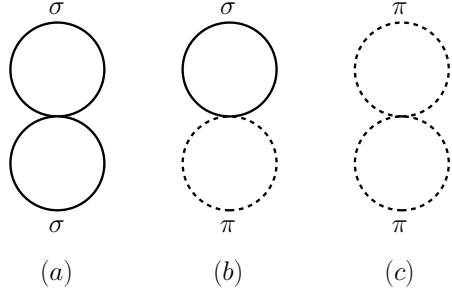}
\caption{\label{Fig:Fig2} The Hartree contributions to the CJT effective potential.
Solid lines correspond to $G_{\sigma}$, while dashed lines
correspond to $G_{\pi}$.}
\end{center}
\end{figure}

Generalising this result to finite isospin chemical potential, we
derive the following expression for the effective potential at
finite temperature
\begin{eqnarray}
V(\phi,G) &=& U(\phi)+\frac{1}{2}\int_{\beta}\ln
G^{-1}_{\sigma}(\phi;k)+\frac{1}{2}\int_{\beta}\ln
G^{-1}_{0}(\phi;k)+\int_{\beta}\ln G^{-1}_c(\phi;k)
\nonumber\\&&+\frac{1}{2}\int_{\beta}[D^{-1}_{\sigma}(\phi;k)G_{\sigma}(\phi;k)-1]
+\frac{1}{2}\int_{\beta}[D^{-1}_{0}(\phi;k)G_{0}(\phi;k)-1]
\nonumber\\&&+\int_{\beta}[D^{-1}_c(\phi;k)G_c(\phi;k)-1]+V_2(\phi,
G_{\sigma}, G_{0}, G_c).
\end{eqnarray}
The first term $U(\phi)$ on the right handside is the classical
potential and the last term $V_2(\phi, G_{\sigma}, G_{0}, G_c)$
denotes the contribution from two--particle irreducible diagrams. In
the following we include only the two--loop diagrams shown in
Fig.~\ref{Fig:Fig2}. These contribute the following terms in the
potential
\begin{eqnarray}
V_2(\phi, G_{\sigma}, G_{\pi^0}, G_c) &=& 3\frac{\lambda}{24}
\left[\int_{\beta}G_{\sigma}^{-1}(\phi;k)
\right]^2+2\frac{\lambda}{24}\left[\int_{\beta}G_{\sigma}^{-1}(\phi;k)\right]\left[\int_{\beta}G_{0}^{-1}(\phi;k)
\right] \nonumber \\&&
+4\frac{\lambda}{24}\left[\int_{\beta}G_{\sigma}^{-1}(\phi;k)\right]\left[\int_{\beta}G^{-1}_c(\phi;k)
\right]+4\frac{\lambda}{24}\left[\int_{\beta}G_{0}^{-1}(\phi;k)\right]\left[\int_{\beta}G_c^{-1}(\phi;k)
\right]\nonumber
\\&&
+3\frac{\lambda}{24} \left[\int_{\beta}G_{0}^{-1}(\phi;k) \right]^2
+8\frac{\lambda}{24} \left[\int_{\beta}G_c^{-1}(\phi;k)\right]^2.
\end{eqnarray}

Minimizing the effective potential with respect to full
propagators we obtain the following system of
nonlinear gap equations:
\begin{subequations}
\begin{eqnarray}
G_{\sigma}^{-1} &=&
D_{\sigma}^{-1}+\frac{\lambda}{2}\int_{\beta}G_{\sigma}(\phi;k)
+\frac{\lambda}{6}\int_{\beta}G_{0}(\phi;k)+\frac{\lambda}{3}\int_{\beta}G_c(\phi;k), \\
G_{0}^{-1} &=&
D_{0}^{-1}+\frac{\lambda}{6}\int_{\beta}G_{\sigma}(\phi;k)
+\frac{\lambda}{2}\int_{\beta}G_{0}(\phi;k)+\frac{\lambda}{3}\int_{\beta}G_c(\phi;k), \\
G^{-1}_c &=& D^{-1}+\frac{\lambda}{6}\int_{\beta}G_{\sigma}(\phi;k)
+\frac{\lambda}{6}\int_{\beta}G_{0}(\phi;k)+\frac{2\lambda}{3}\int_{\beta}G_c(\phi;k).
\end{eqnarray}
\end{subequations}

In order to proceed we can make the following ansatz for the full
propagators:
\begin{eqnarray*}
G_{\sigma}^{-1} &=& \omega_{n}^2+\vec{k}^2+M^2_{\sigma},\\
G_{0}^{-1} &=& \omega_{n}^2+\vec{k}^2+M^2_{0},\\
G^{-1}_c &=& (\omega_{n}+i\mu_{I})^2+\vec{k}^2+M_c^2,
\end{eqnarray*}
where $M_{\sigma}$, $M_{0}$ and $M_c$ are the effective masses of
$\sigma$ meson, $\pi_0$ and charged pion dressed by interaction
contributions from the diagrams of Fig.\ref{Fig:Fig2}. The dressed
sigma and pion masses are then determined by the following gap
equations:
\begin{subequations}\label{e-mass}
\begin{eqnarray}
M_{\sigma}^2 &=&
m^2+\frac{\lambda}{2}\phi^2+\frac{\lambda}{2}F(M_{\sigma})
+\frac{\lambda}{6}F(M_{0})+\frac{\lambda}{3}F(M_c), \\
M_{0}^2 &=&
m^2+\frac{\lambda}{6}\phi^2+\frac{\lambda}{6}F(M_{\sigma})
+\frac{\lambda}{2}F(M_{0})+\frac{\lambda}{3}F(M_c), \\
M_c^2 &=& m^2+\frac{\lambda}{6}\phi^2+\frac{\lambda}{6}F(M_{\sigma})
+\frac{\lambda}{6}F(M_{0})+\frac{2\lambda}{3}F(M_c).
\end{eqnarray}
\end{subequations}
Here we have used a shorthand notation and
introduced the function
\begin{subequations}\label{Eq:FM}
\begin{eqnarray}
F(M_{\sigma})&=& \int_{\beta}\frac{1}{\omega_{n}^2+\vec{k}^2+M^2_{\sigma}},\\
F(M_{0}) &=& \int_{\beta}\frac{1}{\omega_{n}^2+\vec{k}^2+M^2_{0}}, \\
F(M_c) &=&
\int_{\beta}\frac{1}{(\omega_{n}+i\mu_{I})^2+\vec{k}^2+M_c^2}.
\end{eqnarray}
\end{subequations}
At the level of the Hartree  approximation, the thermal effective
masses are independent of momentum and are functions of the order
parameter $\phi$ and the temperature $T$. Moreover, in our case it
is also function of the finite isospin chemical potential $\mu_{I}$.

In terms of the solutions of the gap equations (\ref{e-mass}), the
effective potential at finite temperature can be written in the
form
\begin{eqnarray}\label{Eq:Potential2}
V(\phi,M) &=& \frac{1}{2}m^2\phi^2+\frac{\lambda}{24}\phi^4-\epsilon
\phi+\frac{1}{2}\int_{\beta}\ln
G_{\sigma}^{-1}+\frac{1}{2}\int_{\beta}\ln
G_{0}^{-1}+\int_{\beta}\ln G_c^{-1}\nonumber
\\&&
-\frac{1}{2}(M_{\sigma}^2-m^2-\frac{1}{2}\lambda\phi^2)F(M_{\sigma})
-\frac{1}{2}(M_{0}^2-m^2-\frac{1}{6}\lambda\phi^2)F(M_{0})
\nonumber\\&&
-(M_c^2-m^2-\frac{1}{6}\lambda\phi^2)F(M_c)+\frac{\lambda}{8}[F(M_{\sigma})]^2+\frac{\lambda}{8}[F(M_{0})]^2+\frac{\lambda}{3}[F(M_c)]^2
\nonumber
\\&&
+\frac{\lambda}{12}F(M_{\sigma})F(M_{0})+\frac{\lambda}{6}F(M_{\sigma})F(M_c)+\frac{\lambda}{6}F(M_{0})F(M_c).
\end{eqnarray}
In above expression and in following discussions, we have taken the
symbols $M_{\sigma}$, $M_{0}$ and $M_c$ just as the solutions of the
gap equations (\ref{e-mass}) for simplicity.

Minimizing the effective potential with respect to the full
propagators, we have found the set of nonlinear equations for the
effective particles masses which are given by Eq.~(\ref{e-mass}).
In addition, by minimizing the potential with respect to the order
parameter $\phi$ we obtain one more equation
\begin{eqnarray}\label{phi}
[m^2+\frac{\lambda}{6}\phi^2+\frac{\lambda}{2}F(M_{\sigma})+\frac{\lambda}{6}F(M_{0})+\frac{\lambda}{3}F(M_c)]\phi-\epsilon=0.
\end{eqnarray}

Performing the Matsubara frequency sums as in
Ref.\cite{Dolan:1973qd}, the logarithmic integral which appears in
the above expression for the effective potential
Eq.~(\ref{Eq:Potential2}) divides into two parts. A
zero--temperature part $Q_0(M)$ which is divergent and a
finite--temperature part $Q_{\beta}(M)$ which is finite. For
$S=M_{\sigma}, M_{0}$, we have
\begin{eqnarray}\label{Eq:QM}
Q(S) &=& \frac{1}{2}\int_{\beta}\ln(\omega_{n}^2+\vec{k}^2+S^2) \nonumber\\
&=&
Q_{0}(S)+Q_{\beta}(S) \nonumber \\
&=&
\int\frac{d^3\vec{k}}{(2\pi)^3}\frac{\omega_{\vec{k}}}{2}+\frac{1}{\beta}\int\frac{d^3\vec{k}}{(2\pi)^3}\ln[1-e^{-\beta
\omega_{\vec{k}}}]~,
\end{eqnarray}
where $\omega_{\vec{k}}=\sqrt{(\vec{k}^2+S^2)}$. For $S=M_c$, we get
\begin{eqnarray}\label{Eq:QM2}
Q(S) &=& \int_{\beta}\ln((\omega_{n}+
i\mu_{I})^2+\vec{k}^2+S^2) \nonumber\\
&=&
Q_{0}(S)+Q_{\beta}(S) \nonumber \\
&=& \int\frac{d^3\vec{k}}{(2\pi)^3}\frac{\omega_{\vec{k}}}{2}
+\frac{1}{\beta}\int\frac{d^3\vec{k}}{(2\pi)^3}[\ln(1-e^{-\beta
(\omega_{\vec{k}}+\mu_I)}+\ln(1-e^{-\beta
(\omega_{\vec{k}}-\mu_I)}],
\end{eqnarray}

Similarly, the second type of integral in Eq.~(\ref{Eq:FM}) is
divided into a zero--temperature part $F_{0}(M)$ and a
finite-temperature part $F_{\beta}(M)$. For $S=M_{\sigma}, M_{0}$,
we have
\begin{eqnarray}\label{Eq:FM2}
F(S) &=& \int_{\beta}\frac{1}{\omega_{n}^2+\vec{k}^2+S^2} \nonumber\\
&=& F_{0}(S)+F_{\beta}(S) \nonumber\\
&=& \int\frac{d^3\vec{k}}{(2\pi)^3}\frac{1}{2\omega_{\vec{k}}}
+\int\frac{d^3\vec{k}}{(2\pi)^3}\frac{1}{\omega_{\vec{k}}}\frac{1}{e^{\beta\omega_{\vec{k}}}-1}~.
\end{eqnarray}
And for $S=M_c$, we get
\begin{eqnarray}\label{Eq:FM3}
F(S) &=& \int_{\beta}\frac{1}{(\omega_{n}+i\mu_{I})^2+\vec{k}^2+S^2} \nonumber\\
&=& F_{0}(S)+F_{\beta}(S) \nonumber\\
&=& \int\frac{d^3\vec{k}}{(2\pi)^3}\frac{1}{2\omega_{\vec{k}}}
+\int\frac{d^3\vec{k}}{(2\pi)^3}\frac{1}{2\omega_{\vec{k}}}\left[\frac{1}{e^{\beta(\omega_{\vec{k}}+\mu_I)}-1}
+\frac{1}{e^{\beta(\omega_{\vec{k}}-\mu_I)}-1}\right].
\end{eqnarray}

The second term vanishes at zero temperature while the first term
survives, but the evaluation of the integrals in
Eqs.~(\ref{Eq:QM})(\ref{Eq:QM2})(\ref{Eq:FM2}) and (\ref{Eq:FM3})
requires renormalization. Renormalisation in many--body
approximation schemes is a nontrivial procedure, but in the case of
linear sigma model it has been shown by Rischke and Lenaghan
\cite{Lenaghan:1999si,Lenaghan:2000ey}, that at least in the Hartree
approximation the results are not affected qualitatively. We
therefore simply drop the divergent terms and keep only the finite
temperature part of the integrals. A similar procedure has been
adopted in a number of other investigations using the linear sigma
model \cite{Petropoulos:2004bt}.

Using the above definitions of the integrals in
Eqs.~(\ref{Eq:QM})(\ref{Eq:QM2})(\ref{Eq:FM2}) and (\ref{Eq:FM3}),
we can rewrite the finite temperature effective potential given in
Eq.~(\ref{Eq:Potential2}) in a more compact form
\begin{eqnarray}\label{Eq:Potential3}
V(\phi,M) &=& \frac{1}{2}m^2\phi^2+\frac{\lambda}{24}\phi^4-\epsilon
\phi+Q_{\beta}(M_{\sigma})+Q_{\beta}(M_{0})+Q_{\beta}(M_c)\nonumber
\\&&
-\frac{\lambda}{8}[F_{\beta}(M_{\sigma})]^2-\frac{\lambda}{8}[F_{\beta}(M_{0})]^2
-\frac{\lambda}{3}[F_{\beta}(M_c)]^2-
\frac{\lambda}{12}F_{\beta}(M_{\sigma})F_{\beta}(M_{0})\nonumber
\\&&
-\frac{\lambda}{6}F_{\beta}(M_{\sigma})F_{\beta}(M_c)-\frac{\lambda}{6}F_{\beta}(M_{0})F_{\beta}(M_c).
\end{eqnarray}
This final expression contains only terms which are finite.

\section{\label{Section:Numerical}Numerical results in the Hartree approximation}

\begin{figure}
\includegraphics[scale=0.88]{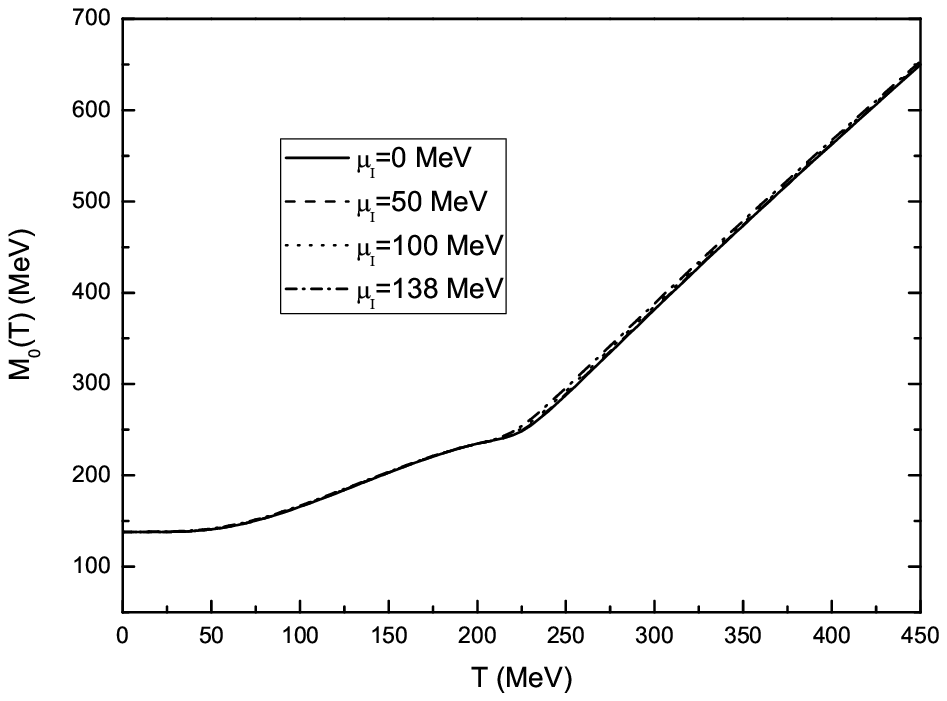}
\caption{Solution of the system of gap equations in the case of
$\epsilon\neq 0$. The effective masse of the neutral pion $M_0$ as a
function of $T$ corresponding to $\mu_{I}=0 \mathrm{MeV}$,
$\mu_{I}=50 \mathrm{MeV}$, $\mu_{I}=100 \mathrm{MeV}$ and
$\mu_{I}=138 \mathrm{MeV}$ respectively.} \label{Fig:Fig3}
\end{figure}

\begin{figure}
\includegraphics[scale=0.88]{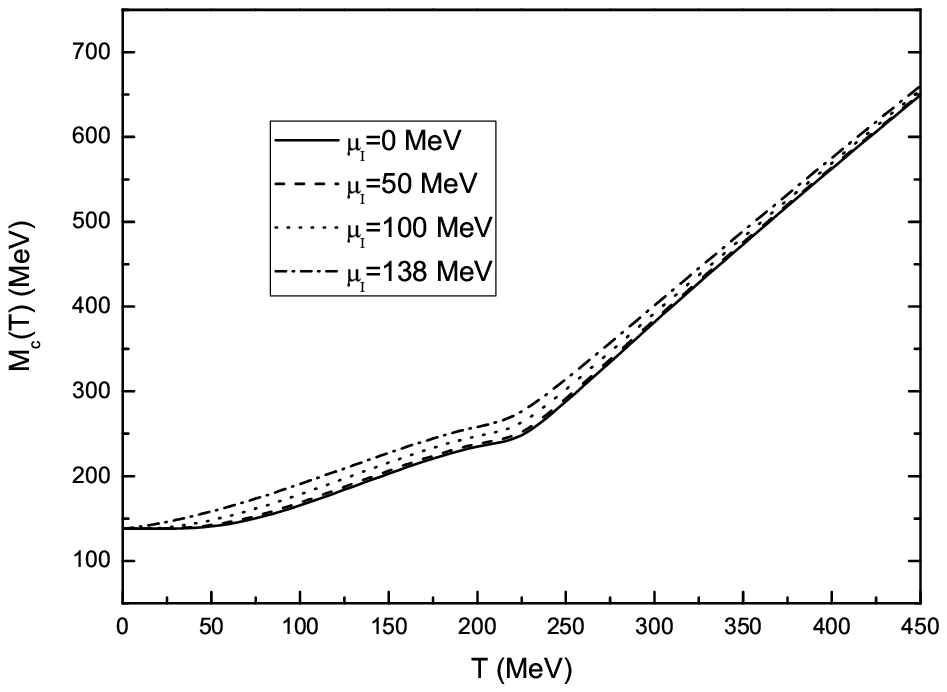}
\caption{Solution of the system of gap equations in the case of
$\epsilon\neq 0$. The effective masse of the charged pion $M_c$ as a
function of $T$ corresponding to $\mu_{I}=0 \mathrm{MeV}$,
$\mu_{I}=50 \mathrm{MeV}$, $\mu_{I}=100 \mathrm{MeV}$ and
$\mu_{I}=138 \mathrm{MeV}$ respectively.} \label{Fig:Fig4}
\end{figure}

\begin{figure}[http]
\includegraphics[scale=0.88]{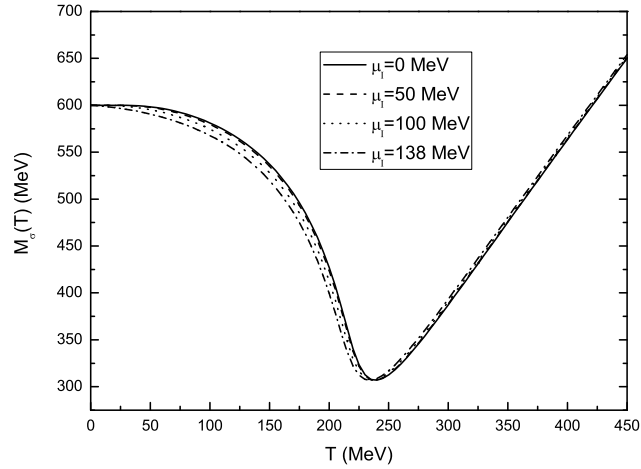}
\caption{\label{Fig:Fig5}Solution of the system of gap equations in
the case of $\epsilon\neq 0$. The effective mass of sigma $M_\sigma$
as a function of $T$ corresponding to $\mu_{I}=0 \mathrm{MeV}$,
$\mu_{I}=50 \mathrm{MeV}$, $\mu_{I}=100 \mathrm{MeV}$ and
$\mu_{I}=138 \mathrm{MeV}$ respectively.}
\end{figure}

We have solved the system of   Eqs.(\ref{e-mass}) and (\ref{phi})
using a numerical method based on the Newton--Raphson method of
solving nonlinear equations. In this way, we are able to determine
the effective masses $M_{\sigma}$, $M_{0}$, $M_c$ and the order
parameter $\phi$ as functions of $\mu_{I}$ and $T$. We are
interested in both negative and positive $\mu_{I}$. The critical
isospin chemical potential at zero temperature is defined by
$\mu_{I}^c=m_{\pi}=138 \mathrm{MeV}$. At $T=0$, for $|\mu_{I}| \leq
m_{\pi}$, the system is in the same ground state as that at
$\mu_{I}=0$. However while it is favorable to have a nontrivial
vacuum which is characterized by the appearance of a condensate
$<\pi^+>$ or $<\pi^->$  when $|\mu_{I}|$ exceeds $m_{\pi}$
\cite{Son:2000xc,Loewe:2002tw,Loewe:2004mu}. Since in this work, we
only consider the effects of the isospin chemical potential on the
masses of mesons and do not discuss the pion condensate
\cite{Son:2000xc,Loewe:2002tw,Loewe:2004mu} or the pion
superfluidity\cite{He:2005nk}, we constrain our discussion to the
values of $\mu_{I}\leq 138 \mathrm{MeV}$.

We show in  Fig.\ref{Fig:Fig3}, Fig.\ref{Fig:Fig4} and
Fig.\ref{Fig:Fig5} the effective masses of the mesons $M_0$, $M_c$
and $M_\sigma$ as functions of $T$ corresponding to $\mu_{I}=0
\mathrm{MeV}$, $\mu_{I}=50 \mathrm{MeV}$, $\mu_{I}=100 \mathrm{MeV}$
and $\mu_{I}=138 \mathrm{MeV}$ respectively. For $\mu_{I}=0
\mathrm{MeV}$, at finite $T$ we reproduce the well known results for
the effective masses $M_0$, $M_c$ and $M_{\sigma}$, i.e., see
Refs\cite{Petropoulos:2004bt}. For $\mu_{I}\neq 0 \mathrm{MeV}$, the
behaviors of the meson masses at finite $T$ are very similar to the
case of $\mu_{I}=0 \mathrm{MeV}$. At low temperature, the pions
appear with the observed mass, but their masses increase with
temperature while the sigma mass decreases with temperature. At high
temperature(higher than $\sim 300$ MeV), due to interactions in the
thermal bath, the interactions in the thermal bath, all particles
(pions and sigma) appear to have the same effective mass. At $T=0
\mathrm{MeV}$, the masses of the pions and sigma particle remain
constant for different $\mu_{I}$ .

In Fig.\ref{Fig:Fig3}, the results for neutral pions at various
finite isospin chemical potential are nearly identical to each
other. This means that the finite isospin chemical potential
$\mu_{I}$ almost can not affect the effective masses of these
particle duo to the fact that the neutral pions do not contribute to
the finite isospin chemical potential $\mu_{I}$. However, from
Eq.(\ref{eq-pro3}), for a non-vanishing isospin chemical potential,
the masses of the charged pions at the tree level will acquire a
nontrivial isospin chemical potential dependence for finite values
of temperature. In Fig.\ref{Fig:Fig4}, as the temperature increases,
the effective mass of the charged pions at finite temperature
changes  considerably for different values of the isospin chemical
potential, especially at the range of  $T=50 \sim 300 MeV$.
Moreover, from Fig.\ref{Fig:Fig3} and Fig.\ref{Fig:Fig4}, for a
certain temperature region, the masses of pions (both charged pion
and neutral pion)increase with the increase of the isospin chemical
potential $\mu_{I}$, though the mass of the neutral pion changes
very slightly.

In Fig.\ref{Fig:Fig5}, the results for sigma particle are very
difference from the poins. When the temperature is $T< 235
\mathrm{MeV}$, for a fixed temperature, the mass of sigma decreases
considerably with the increase of the isospin chemical potential
$\mu_{I}$, while for temperature $T> 235 \mathrm{MeV}$, for fixed
temperature, the mass of sigma increases very slightly increases  as
 isospin chemical potential $\mu_I$ increases. In this case, the results for the sigma particle mass
are nearly identical to the ones for the neutral pions.

\begin{figure}[http]
\includegraphics[scale=0.9]{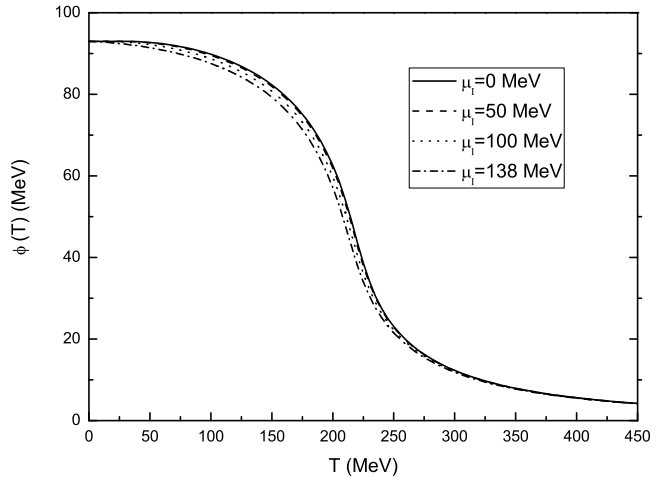}
\caption{\label{Fig:Fig6} Evolution of the order parameter $\phi$ as
a function of $T$ corresponding to $\mu_{I}=0 \mathrm{MeV}$,
$\mu_{I}=50 \mathrm{MeV}$, $\mu_{I}=100 \mathrm{MeV}$ and
$\mu_{I}=138 \mathrm{MeV}$ respectively.}
\end{figure}

The presence of the finite isospin chemical potential $\mu_{I}$
modifies the evolution of the order parameter $\phi$. As it is shown
in Fig.\ref{Fig:Fig6}, at low temperature, for a fixed temperature,
the order parameter decreases considerably as the isospin chemical
potential increases. But at very high temperature, the mass of sigma
is very slightly changed with the increase of the isospin chemical
potential $\mu_{I}$. For various $\mu_{I}$, when the temperature
increases, the order parameter decreases, and vanishes smoothly at
very high temperatures. In this case, the change is not a phase
transition any more and we encounter a smooth crossover from a low
temperature phase to a high temperature phase. We conclude that the
presence of a finite isospin chemical potential when $|\mu_{I}| \leq
m_{\pi}$ does not change the nature of the chiral phase transition,
only modifies the effective masses of mesons in the linear sigma
model, in accordance with earlier studies
\cite{Roder:2003uz}\cite{Lenaghan:1999si}.

\section{\label{Section:Conclusions} Conclusions}

In summary, we have discussed the effect of the finite isospin
chemical potential to the effective masses of the measons at finite
temperature in the framework of the $O(4)$ linear sigma model with
explicit chiral symmetry breaking. We have provided an alternative
mechanism to include the isospin chemical potential in the linear
sigma model. In order to get some insight into how the the isospin
chemical potential could affect the chiral phase transition, we have
solved the system of gap euqations numerically and found the
evolution with temperature of the effective thermal masses of mesons
at various $\mu_{I}$.

For neutral pions, the results for different values of finite
isospin chemical potential are nearly identical to each other due to
the fact that the neutral pions do not contribute to the finite
isospin chemical potential $\mu_{I}$. But for charged pions, the
situation is very different. At low temperature, the effective
masses of charge pions are very different than that of neutral pions
due to the effects of the finite isospin chemical potential, it
appears that they increase more dramatically than that of neutral
pions as the temperature increasing. But for high temperature, the
situation has been changed, the effective mass of charged pions are
similar to that of neutral pions. On the contray, the results for
sigma particle are very difference from the poins. For a fixed
temperature, when $T< 235 \mathrm{MeV}$, the mass of sigma decreases
considerably with the increasing $\mu_{I}$, while for $T> 235
\mathrm{MeV}$, the mass of sigma  increases very slightly as $\mu_I$
increases as well. The evolution of the order parameter $\phi$ for
different $\mu_{I}$ shows the effects of the finite isospin chemical
potential do not change the nature of chiral phase transition, but
just modify the effective masses of mesons in the linear sigma
model.

Of course, as we have already pointed out, we have not considered
the pion condensate  \cite{Son:2000xc,Loewe:2002tw,Loewe:2004mu} or
the pion superfluidity\cite{He:2005nk}. Our discussions for are
restricted the values of $\mu_{I}\leq 138 \mathrm{MeV}$. Since there
will be more complicated and richer vacuum structures when $\mu_{I}$
is larger than $138\mathrm{MeV}$, it is of interest to extend our
work to finite temperature for $\mu_{I}> 138 \mathrm{MeV}$ and
discuss the pion condensate and pion superfluidity. According to the
work of Shu and Li, our work can be used to study the Bose-Einstein
condensation and chiral phase transition with explicit chiral
symmetry breaking. All these works are in progress.

\begin{acknowledgments}
The authors wish to thank Tao Huang, Xinmin Zhang and especially
Pengfei Zhuang for useful discussions. This work is supported in
part by the National Natural Science Foundation of People's Republic
of China. NP would like to thank the members of the theory Group of
the University of Coimbra for their warm hospitality, since part of
this work was carried out there, and especially Professor Eef van
Beveren.
\end{acknowledgments}

\end{document}